\documentclass[journal]{IEEEtran}

\usepackage[pdftex]{graphicx}
\graphicspath{{../pdf/}{../jpeg/}}
\DeclareGraphicsExtensions{.pdf,.jpeg,.png}

\usepackage[cmex10]{amsmath}
\usepackage{mdwmath}
\usepackage{mdwtab}
\usepackage{eqparbox}
\usepackage{url}

\usepackage{tikz,xcolor,hyperref}
\definecolor{lime}{HTML}{A6CE39}
\DeclareRobustCommand{\orcidicon}{%
    \begin{tikzpicture}
    \draw[lime, fill=lime] (0,0) 
    circle [radius=0.16] 
    node[white] {{\fontfamily{qag}\selectfont \tiny ID}};    \draw[white, fill=white] (-0.0625,0.095) 
    circle [radius=0.007];    \end{tikzpicture}
    \hspace{-2mm}}
\foreach \x in {A, ..., Z}{%
    \expandafter\xdef\csname orcid\x\endcsname{\noexpand\href{https://orcid.org/\csname orcidauthor\x\endcsname}{\noexpand\orcidicon}}
    }

\ifCLASSINFOpdf
\else
\fi
\begin{document}
%
\title{Radar-STDA: A High-Performance Spatial-Temporal Denoising Autoencoder for Interference Mitigation of FMCW Radars}
%
%
%

\author{Lulu Liu,
        Runwei Guan\orcidA{},~\IEEEmembership{Student Member,~IEEE},
        Fei Ma\orcidD{},
        Ka Lok Man\orcidC{},
        Jeremy Smith,~\IEEEmembership{Member,~IEEE},
        Yutao Yue\orcidB{}
        
\thanks{Lulu Liu and Fei Ma are with School of Science, Xi'an Jiaotong-Liverpool University, Suzhou 215123, China (email: Lulu.Liu21@student.xjtlu.edu.cn; Fei.Ma@xjtlu.edu.cn).}

\thanks{Runwei Guan, Ka Lok Man and Jeremy Smith are with School of Advanced Technology, Xi'an Jiaotong-Liverpool University, Suzhou 215123, China (email: Runwei.Guan21@student.xjtlu.edu.cn; Ka.Man@xjtlu.edu.cn).}

\thanks{Lulu Liu is with School of Physical Sciences, University of Liverpool, Liverpool L69 3GJ, United Kingdom (email: Lulu.Liu21@student.xjtlu.edu.cn).}

\thanks{Runwei Guan is with Department of Electrical Engineering and Electronics, University of Liverpool, Liverpool L69 3GJ, United Kingdom (email: runwei.guan@liverpool.ac.uk).}

\thanks{Lulu Liu, Runwei Guan and Yutao Yue are with XJTLU-JITRI Academy of Industrial Technology, Xi'an Jiaotong-Liverpool University, Suzhou 215123, China (email: Lulu.Liu21@student.xjtlu.edu.cn; Runwei.Guan21@student.xjtlu.edu.cn; yueyutao@idpt.org).}

\thanks{Yutao Yue is with Department of Mathematical Sciences, University of Liverpool, Liverpool L69 7ZX, United Kingdom, and also with Institute of Deep Perception Technology, JITRI, Wuxi 214000, China (email: yueyutao@idpt.org).}
}
%
%

\markboth{Journal of \LaTeX\ Class Files,~Vol.~14, No.~8, August~2015}%
{Shell \MakeLowercase{\textit{et al.}}: Bare Demo of IEEEtran.cls for IEEE Journals}
%



\maketitle

\begin{abstract}
With its small size, low cost and all-weather operation, millimeter-wave radar can accurately measure the distance, azimuth and radial velocity of a target compared to other traffic sensors. However, in practice, millimeter-wave radars are plagued by various interferences, leading to a drop in target detection accuracy or even failure to detect targets. This is undesirable in autonomous vehicles and traffic surveillance, as it is likely to threaten human life and cause property damage. Therefore, interference mitigation is of great significance for millimeter-wave radar-based target detection. Currently, the development of deep learning is rapid, but existing deep learning-based interference mitigation models still have great limitations in terms of model size and inference speed. For these reasons, we propose Radar-STDA, a Radar-Spatial Temporal Denoising Autoencoder. Radar-STDA is an efficient nano-level denoising autoencoder that takes into account both spatial and temporal information of range-Doppler maps. Among other methods, it achieves a maximum SINR of 17.08 dB with only 140,000 parameters. It obtains 207.6 FPS on an RTX A4000 GPU and 56.8 FPS on an NVIDIA Jetson AGXXavier respectively when denoising range-Doppler maps for three consecutive frames. Moreover, we release a synthetic data set called Ra-inf for the task, which involves 384,769 range-Doppler maps with various clutters from objects of no interest and receiver noise in realistic scenarios. To the best of our knowledge, Ra-inf is the first synthetic dataset of radar interference. To support the community, our research is open-source via the link \url{https://github.com/GuanRunwei/rd_map_temporal_spatial_denoising_autoencoder}.

\end{abstract}


%
\IEEEpeerreviewmaketitle

\section{Introduction}
%
%
%
%

\IEEEPARstart{M}{illimeter}-wave radar is a significant sensor in the field of transportation, where one of the most commonly used modulations is frequency-modulated continuous wave (FMCW). It could detect the range, azimuth, velocity and radar cross-section (RCS) of the target. Radar has a strong ability to penetrate dust and smoke. Compared with cameras and LIDAR, radar is more robust in dark environments and adverse weather \cite{bai2021robust}, and its cost is relatively low. In addition, the small size of radar makes it easy to install and deploy. With the rapid development of artificial intelligence, it could make sensors smarter in autonomous vehicles and intelligent transportation systems. For autonomous vehicles, high detection sensitivity is a must to guarantee personal and property safety in kinds of contexts, e.g., adaptive cruise control and automatic emergency braking. For intelligent transportation systems, perceiving traffic flow precisely could help traffic supervisors make decisions more reasonable. Nevertheless, there is the potential to induce mutual interference with the rapidly rising number of automotive radar sensors and the limited bandwidth. Interference can therefore cause severe degradation for sensors in reality as well as noise and clutters. 

There are a number of ways to avoid interference, such as standardization, ‘sense and avoid’ and polarization. In many cases, however, it is impossible to avoid mutual interferers completely. In such situations, an approach of localization followed by mitigation is used to reduce degradation, such as Zeroing \cite{fuchs2021complex}, Iterative method with adaptive thresholding (IMAT) \cite{bechter2017automotive} and Ramp filtering \cite{wagner2018threshold}. Afterward, more sophisticated algorithms are exploited, including the computation time reduced compressive sensing \cite{chen2022iterative}, the time-domain low-pass filter with wavelet denoising \cite{lee2019mutual}\cite{xu2021interference} and the CFAR-based interference mitigation \cite{wang2021cfar}. By analyzing and exploring the aforementioned methods, however, we have identified a number of drawbacks and limitations to traditional interference mitigation methods based on signal processing. Firstly, signal-processing-based methods are normally two-stage, consisting of interference detection and interference removal. The performance of removal heavily relies on the method of interference detection. Secondly, they lack adaptivity and rely on manual hyperparameters when confronting various interference. Thirdly, in some circumstances, signal-processing-based methods remove part of useful information, which causes the distortion of the target peak.

Deep-learning-based (DL-based) methods do not make prior assumptions on the data of noise and interference, which could directly process the data and maintain the most useful signal, outperforming the traditional methods under severe interference conditions. For example, Jiwoo Mun et.al \cite{mun2020automotive} introduced the self-attention mechanism in a three-layer Recurrent Neural Network (RNN) and restored the beat signal in the time domain. However, there is no frequency domain information used in the network. In \cite{ristea2020fully}, the researchers performed a discrete Short-Time Fourier Transform (STFT) to obtain time-frequency spectrograms, which are then trained by a fully convolutional neural network. They also released a large-scale simulated data set as open source in \cite{ristea2021estimating}. Rock and Fuchs \cite{fuchs2021complex} proposed the use of Complex-Valued Convolutional Neural Networks (CVCNNs) to address the issue of phase loss. Furthermore, they propose a quantized convolutional neural network to reduce the memory occupation rate and computational complexity \cite{rock2021resource}, enabling real-time processing on edge devices. However, firstly, most current DL-based methods adopt simple fully convolutional networks as the denoising autoencoders. Considering the weight-sharing nature of convolution, it makes little sense that many features of noise and interference are still retained in the denoised results. Secondly, the signal-to-noise ratio of range-Doppler (RD) map is too low to restore the target peak in some circumstances. Thirdly, few training data would lead to over-fitting of the autoencoder and the recovered range-Doppler map would differ significantly from the ground truth. Last but not least, almost all works do not consider the real-time performance and utility of models in different devices. Hence, interference suppression remains a huge challenge for industry and research nowadays. To address the problems, we propose an efficient denoising autoencoder named Radar-STDA for interference mitigation of range-Doppler maps. Moreover, we release a synthetic dataset called Ra-inf for interference mitigation combined with RaDICaL \cite{lim2021radical}. Our contributions are as follows, 


\begin{enumerate}
    \item We release a data set called Ra-inf synthesized with real-world data for interference mitigation, which is close to realistic measurements, thus possessing better performance of generalization compared to purely simulated data sets.
    \item We analyze the effects of FMCW interference parameters on victim signals via simulation. It shows that various patterns can be formed according to the relative value of a few parameters, e.g., sweep duration, sweep bandwidth and carrier frequency.
    \item We propose a nano-level but highly effective denoising autoencoder called Radar-STDA. In Radar-STDA, we design a lightweight encoder (decoder) architecture called mobile encoder (decoder). Radar-STDA adaptively fuses both spatial and temporal information with channel attention, and gets 17.08 dB signal-to-interference-and-noise-ratio (SINR). It outperforms other signal-processing-based models and normal DL-based models.
    \item Radar-STDA has only 0.14 million parameters, which is far fewer than any other denoising autoencoders and easily deployed. It gets 207.6 FPS on one RTX A4000 GPU and 56.8 FPS on NVIDIA Jetson AGX Xavier when denoising  range-Doppler maps of three consecutive frames.
\end{enumerate}

\section{Related Works}






For the rapid development of deep learning, denoising autoencoder is applied for interference mitigation of automotive radar. \cite{rock2019cnns} adopted a normal convolutional neural network to mitigate noises on range-Doppler maps. \cite{rock2019complex} adopted a normal convolutional neural network to denoise range-Profile maps and range-Doppler maps. \cite{de2020deep} proposed CAE, introducing spatial and temporal information at the same time for noise reduction, which could replace CFAR and peak detection operation. \cite{abdulatif2019towards} designed a UNet-based \cite{ronneberger2015u} neural network for the denoising and reconstruction of micro-Doppler maps based on the generative adversarial network.


Furthermore, fast inference on edge devices is significant in many traffic scenarios. Neural networks (NNs) that meet the requirements usually have few parameters. MobileNet V1 \cite{howard2017mobilenets} adopted depthwise separable convolution to reduce the parameter number dramatically. MobileNet V2 \cite{sandler2018mobilenetv2} used inverted residual blocks with linear activation. MobileNet V3 \cite{howard2019searching} introduced the neural architecture search to the network and a lightweight attention module called squeeze and excitation (SE). ShuffleNet \cite{zhang2018shufflenet} adopted pointwise group convolution and channel shuffle, which could greatly reduce the computational burden of the neural network while maintaining accuracy.

\section{Automotive FMCW Radar Signal Model and Interference Analysis}

The fast chirp FMCW modulation as a variant of the classical FMCW modulation is commonly used in commercial automotive radar systems, due to its simple signal dechirping procedure and relatively low requirements of analog-to-digital converter (ADC). Mutual interference happens with the increasing number of millimeter wave radar-equipped vehicles. In radar signal processing, Range-Doppler response is usually computed to conduct object detection and further operation in the frequency domain. It's nonnegligible that the stronger interference compared with objects under the same conditions can induce large effects, such as false alarm or missing detection. Therefore, interference mitigation has become a significant task.

\subsection {Radar Signal Model}

\begin{figure}[ht!]
\centering
\includegraphics[width=3.4in]{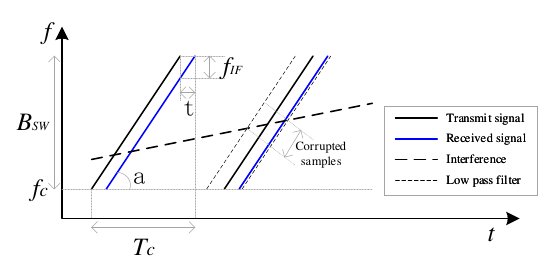}
\caption{Illustration of the FMCW radar principle..}
\label{radar principle}
\end{figure}

In typical fast chirp FMCW radar systems \cite{kronauge2014new}, a radar sensor transmits a sequence of linear frequency chirps, also termed ramps. The frequency of the chirps follows a sawtooth pattern, as depicted in Fig.~\ref{radar principle}. The frequency of each chirp signal with the carrier frequency $f_c$, sweep bandwidth $B_{SW}$ and sweep duration $T_c$ is expressed as

\begin{equation}\label{FMCW chirp frequency}
    f(t)=f_c+\frac{B_{SW}}{T_c}t,
\end{equation}
where $0 \leq t < T_c$ denotes the time variable in the fast-time dimension, and $\frac{B_{SW}}{T_c}$ can be written as the chirp rate $\alpha$. The corresponding transmitted signal with $M$ repeated chirps can be written as

\begin{flalign*} 
\label{chirp signal}
    x_T(t,m)=&A_T\exp{(j2\pi\int_0^t f(t)\mathrm{d}t)}\\
    =&A_T\exp{(j2\pi(f_c(t-mT_c)}\\
    &{+\frac{B_{SW}}{2T_c}(t-mT_c)^2)+j\phi)},
\end{flalign*}    

where $A_T$ is the amplitude of the transmitted signal, $0 \leq m < M$ denotes the chirp index in the slow-time dimension, and $\phi$ is the initial phase of the signal. Since the round-trip time delay $\tau$ at the receiver is caused by the target reflection, the $m$th received chirp signal for a single target is
 
\begin{equation}\label{receive signal}
    x_R(t,m)=A_Rx_T(t-mT_c-\tau),
\end{equation}
where $A_R$ denotes the received amplitude and $\tau=2(D+vt)/c$. Here, $D$ and $v$ represent the distance and relative radial velocity between the target and the sensor, respectively, and $c$ is the speed of light. Subsequently, the beat signal at the intermediate frequency (IF) is obtained by mixing the conjugate of $x_R(t,m)$ with the transmitted signal $x_T(t,m)$, computed as:

\begin{equation}\label{beat signal}
    s_M(t,m)=x_R^\ast(t,m)x_T(t,m).
\end{equation}

Following a low-pass filter (LPS) and an ADC with a sampling frequency of $1/T_s$ and $N$ samples per chirp, the discrete beat signal can be approximated as \cite{engels2021automotive}

\begin{equation}\label{discrete signal}
\begin{aligned}
    s_B(n,m)\approx&{A_RA_T\exp{(j2\pi\frac{B_{SW}}{T_c}\frac{2D}{c}nT_s)}}\\
    &\cdot\exp{(j2\pi{f_c}\frac{2v}{c}mT_c)}\cdot\exp{(j2\pi{f_c}\frac{2D}{c})}\\
    &+v(n,m),
\end{aligned}    
\end{equation}
where $0 \leq n < N$, and $\frac{2v}{c} f_C$ can be denoted as the Doppler shift $f_D$ that is resulted from the target motion. Considering that the mutual interference signal is linearly superimposed on the multiple reflections, the corrupted signal is

\begin{equation}\label{IF signal}
        s(n,m)=\sum_{o=1}^{N_O}{s_{B,o}(n,m)}+\sum_{i=1}^{N_I}{s_{I,i}(n,m)}+v(n,m),
\end{equation}
where $N_O$ and $N_I$ correspond to the number of targets and the number of interfering sensors, separately, and $v(n,m)$ is the receiver thermal noise approximated by complex-valued Gaussian white noise. $s_{I,i}(n,m)$ is sampled from the $i$th interference signal $s_{I,t}(t)$ which is defined as 

\begin{equation}
    s_{I,i}(t)=x_{Int,i}(t-\Tilde{\tau_i})x_T(t).
\end{equation}
where $x_{Int,i}(t-\Tilde{\tau_i})$ is the $i$th interfering signal with the time delay $\Tilde{\tau_i}$.

\subsection {Range-Doppler Processing}

Consider again the IF signal $s(n,m)$ in (\ref{IF signal}). The set of $\left\{s(n,m), 0 \leq n <N, 0 \leq m <M\right\}$ composes an $N \times M$ data matrix. Thus two-dimensional Fast Fourier Transform (2D-FFT) can be performed firstly on the matrix in the fast-time dimension, followed by a row-wise Fourier transform in the slow-time dimension. The outcome is referred to as range-Doppler (RD) map as followings,

\begin{equation}\label{column transform}
    y(p,m)=\sum_{n=1}^{N}{(s(n,m)\exp{(-j\frac{2\pi n}{P}p)})},
\end{equation}
\begin{equation}\label{row transform}
    y_{RD}(p,q)=\sum_{m=1}^{M}{(y(p,m)\exp{(-j\frac{2\pi m}{Q}q)})}.
\end{equation}
Here, $0 \leq p <P$ and $0 \leq q < Q$. $P$ and $Q$ are the number of points in the column-wise Fourier transform and the number of points in the horizontal transform. According to (\ref{column transform}), we can detect a peak at the frequency of $\frac{2B_{SW}D}{cT_c}$ in (\ref{discrete signal}) which implies the radial distance between the target and the radar sensor. After applying Fourier transform across the slow-time dimension as computed in (\ref{row transform}), peaks that account for the relative radial velocity can be obtained at the frequency of $\frac{2f_cv}{c}$. Therefore, we can simultaneously estimate both the range and velocity of each target on RD maps \cite{kim2020classification}.

\subsection {Power level of reflection and interference}

In general, interference power is much stronger than the desired signal reflected from targets, since the interference wave is one-way propagated while reflections travel two ways. The received interference power in the Friis free space and the radar equation are

\begin{equation}
    P_{int}=P_T\frac{G_{trx}\lambda^2}{(4\pi)^2r^2},
\end{equation}
\begin{equation}
    P=P_T\frac{G_{trx}\sigma\lambda^2}{(4\pi)^3d^4},
\end{equation}
respectively, where $r$ is the distance between the interferer and the victim radar, $d$ is the distance between the radar and the target, $P_T$ is the transmit power, $G_{trx}$ is the combined transmit and receive antenna gain, $\sigma$ is the RCS of the target \cite{aydogdu2019radar}. Hence, when $r$ and $d$ are similar, the interference power exceeds the signal power with a typical RCS value, that is $P_{int} \gg P$. Besides, the effect of interference depends on the aggregate power of the interference samples, and the extent of coherence between victim and aggressor radars \cite{goppelt2010automotive}.

\subsection{Analysis for the impact of interference}

We consider only FMCW radar interfering with FMCW radar here. To validate our analysis, we show various simulation results based on the radar system parameters in Table \ref{tab:radar parameters}. Victim signal and interference signal are generated via MATLAB Radar Toolbox. For FMCW radar, the sweep bandwidth, the sweep duration, and the carrier frequency are adjusted according to Table \ref{tab:FMCW parameters}, in which the numbers in the middle two rows represent the ratio between the ramps of the interfering radar and the victim radar. For example, $1.1=B_{SWint}/B_{SWvic}$. The numbers in the last column mean the difference between the carrier frequencies of the interference and victim ramps.

\begin{table}[]
    \centering
    \caption{Generic radar system parameters of simulated victim and aggressor}
    \begin{tabular}{ccc}
         \hline
         Parameters & Victim & Aggressor \\
         \hline
         Carrier frequency [GHz] & 77 & 77 \\
         Transmit power [dBm] & 5 & 5 \\
         Transmit antenna gain [dB] & 36 & 36 \\
         Receive antenna gain [dB] & 42 & - \\
         Receiver noise figure [dB] & 4.5 & - \\
         \hline
    \end{tabular}    
    \label{tab:radar parameters}
\end{table}

\begin{table}[]
    \centering
    \caption{Quanlitative interference analysis of seven typical scenarios}
    \begin{tabular}{cccccccc}
    \hline
         Scenario & (1) & (2) & (3) & (4) & (5) & (6) & (7)\\
         \hline
         Ratio of $B_{SW}$ & 1 & 1 & 1 & 2 & 1 & 2 & 2\\
         Ratio of $T_c$ & 1 & 1.1 & 2 & 1 & 1 & 1.1 & 2\\
         $f_c$ offset (MHz) & 0 & 0 & 0 & 0 & -20 & 0 & 0\\
         \hline
    \end{tabular}
    \label{tab:FMCW parameters}
\end{table}

\begin{figure}[ht!]
\centering
\includegraphics[width=3.4in]{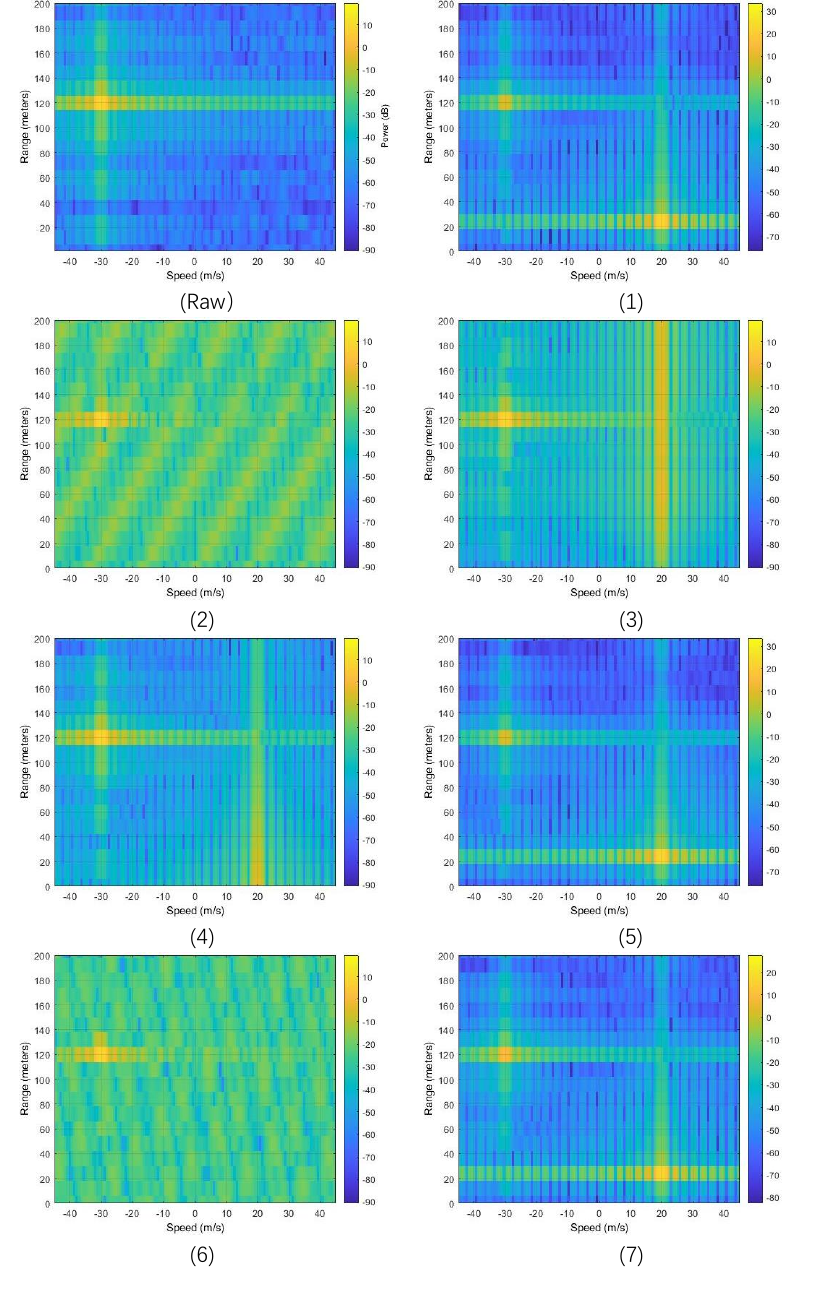}
\caption{Range Doppler plot of varying interference phenomena}
\label{RD maps of interference}
\end{figure}

Fig. \ref{RD maps of interference} illustrates the RD maps for typical interference scenarios. In the simulation, the raw RD map having no interference consists of one target on the left. As we can see from scenarios (1) and (7), the interference appears as a target when the parameters are identical or integer multiple, termed ghost. Scenario (5) explains that adjusting the carrier frequency of interference rarely influences the received signal. As the sweep duration of interference differs slightly in (2) and (6), the interference power is distributed throughout the whole spectrum, resulting in the increased noise floor. In scenario (3), there's a 'ridge' on the RD map because of the integer multiple of the sweep duration. A similar phenomenon occurs when the sweep bandwidth is adjusted as given in (4). Consequently, missing targets or false alarms can occur under the above circumstances, and it's indispensable to suppress interference.

\section{Radar-STDA}

Radar-STDA is an efficient lightweight denoising autoencoder for interference mitigation of radar, whose full name is Radar-Spatial Temporal Denoising Autoencoder. It is for the interference mitigation and range-Doppler map restoration of victim radars. The features and advantages of Radar-STDA are as follows:

\begin{enumerate}
    \item It fuses the spatial and temporal information of range-Doppler maps under the online pattern to mitigate interference of range-Doppler maps.
    \item Efficient channel attention (ECA) is adopted tactfully to mitigate interference adaptively.
    \item It is a nano-level denoising autoencoder with fast inference speed, which is friendly deployed and run on edge devices.
\end{enumerate}

\subsection{The Overall Architecture of Radar-STDA}

\begin{figure*}
\centering
\includegraphics[width=0.95\textwidth]{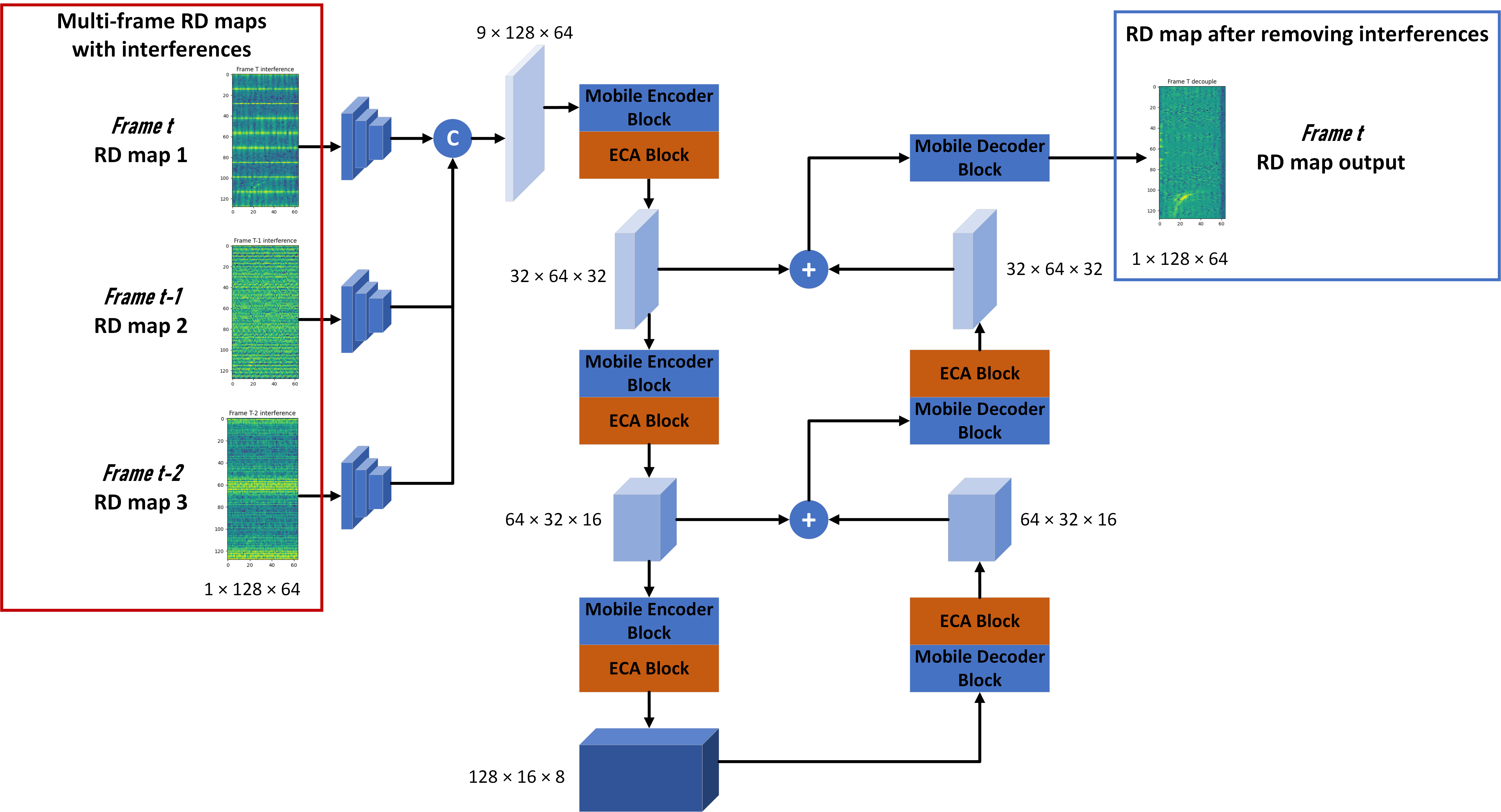}
\caption{The architecture of Radar-STDA. The shape of range-Doppler maps in our experiments are all $1 \times 128 \times 64$.}
\label{radar_stda_architecture}
\end{figure*}

As Fig. \ref{radar_stda_architecture} presents, there are four modules in Radar-STDA, which are the input, encoder, decoder and output. 

Radar-STDA takes three range-Dopper maps of three adjacent frames as the input of the encoder $h(\cdot)$, which is shown in Eq. \ref{encoder_stda}.

\begin{equation}
    y = h(x_t, x_{t-1}, x_{t-2}),
\label{encoder_stda}
\end{equation}
where $y$ is the output of the encoder $h(\cdot)$. $\{x_t, x_{t-1}, x_{t-2}\}$ are the input of the encoder $h(\cdot)$, which are range-Doppler maps at frame $t$, $t-1$ and $t-2$ with interferences. The encoder transforms the feature maps from a low-dimensional space to a high-dimensional space.

The decoder $f(\cdot)$ is to remove interferences and restore the original range-Doppler map at frame $t$ as much as possible. The whole process is presented in Eq. \ref{decoder_stda}.

\begin{equation}
   z = f(y) = f(h(x_t, x_{t-1}, x_{t-2})),
\label{decoder_stda}
\end{equation}
where $z$ is the restored range-Doppler map at frame $t$ after removing interferences.

\subsection{Fusion of Spatial-Temporal Information}

\begin{figure}
\centering
\includegraphics[width=0.47\textwidth]{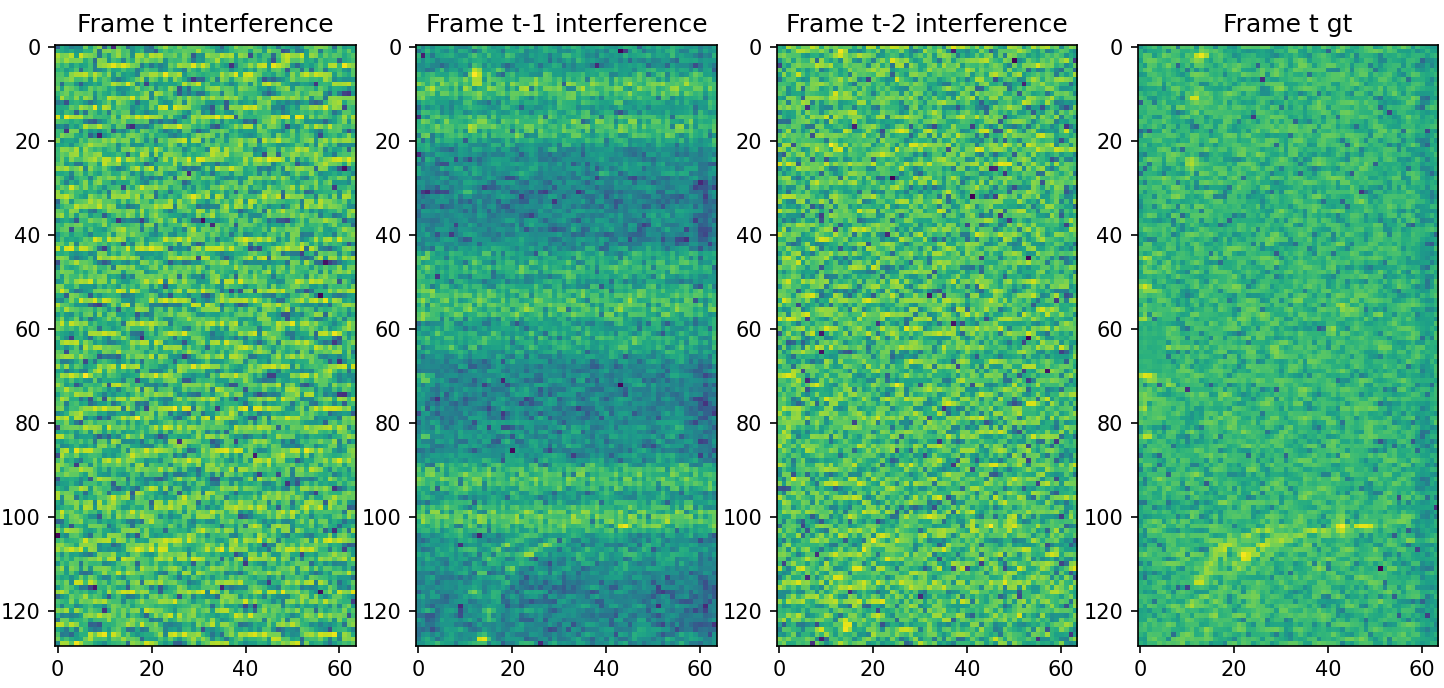}
\caption{A batch of three range-Doppler maps with interferences at the three temporal adjacent frames. \textbf{Frame t gt} is the ground-truth of the range-Doppler map at frame $t$, which is the original range-Doppler map without interference.}
\label{fusion_reason}
\end{figure}

Aside from the conventional denoising autoencoder, our Radar-STDA takes range-Doppler maps at the three temporal adjacent frames as the input, which includes the range-Doppler map of the current frame and the previous two frames. It is an online denoising pattern that Radar-STDA could run in real time. As Fig. \ref{fusion_reason} shows, in some cases, the Signal to Interference and Noise Ratio (SINR) of range-Doppler is low. Intuitively, the target in the range-Doppler map at frame $t$ could not be recognized. Therefore, we need the range-Doppler maps at previous frames to assist the restoration of the map at frame $t$. More exactly, Radar-STDA concatenates three range-Doppler maps from the dimension of the channel. Assuming the input maps are $x_{t}, x_{t-1}, x_{t-2} \in R^{c \times h \times w}$, the concatenation operation is shown in Eq. \ref{concat_operation}.

\begin{equation}
    x_{st} = Concat(x_{t}, x_{t-1}, x_{t-2}), x_{st} \in R^{3c \times h \times w}.
\label{concat_operation}
\end{equation}

From another perspective, multi-frame information means that there are various interferences in the features of range-Doppler maps, because there may be different interferences in different frames. Therefore, in the training process, Radar-STDA needs to remove various interferences as much as possible, which indirectly enhances the capability of Radar-STDA for interference mitigation and map restoration.

\subsection{Adaptive Mobile Encoder-decoder Module with Efficient Channel Attention (ECA)}

\begin{figure}
\centering
\includegraphics[width=0.47\textwidth]{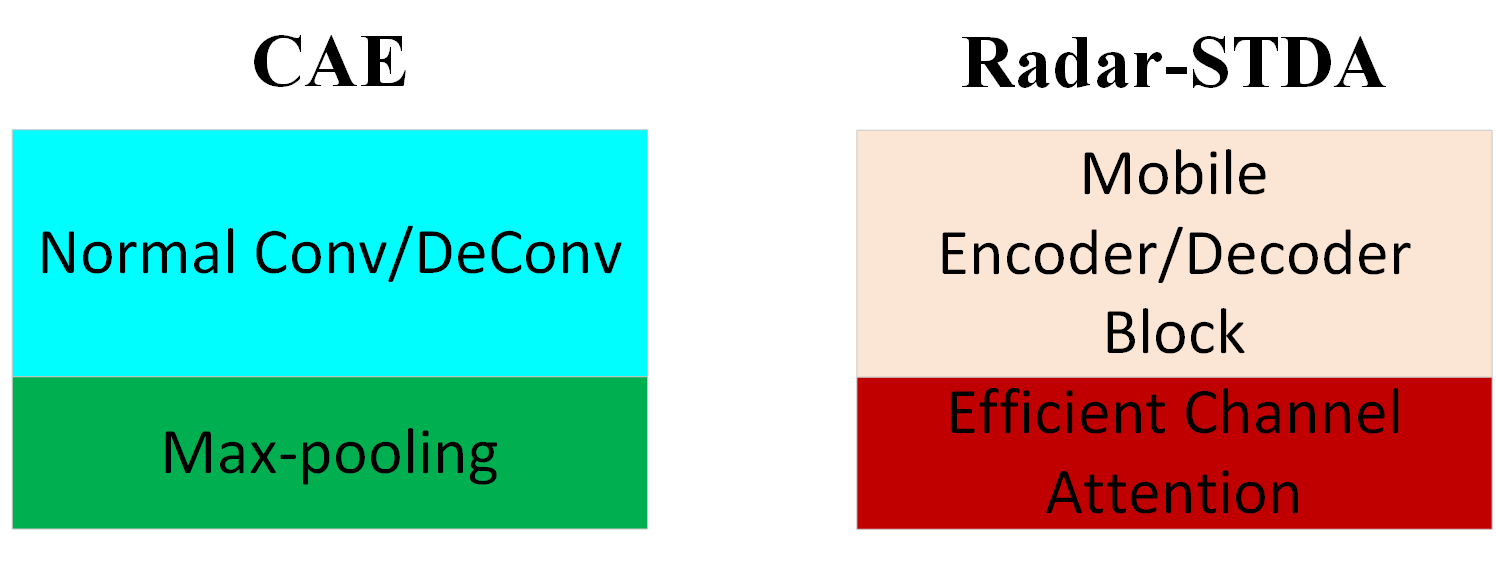}
\caption{The comparison between the units of CAE and Radar-STDA.}
\label{encoder_decoder}
\end{figure}

\begin{figure}
\centering
\includegraphics[width=0.47\textwidth]{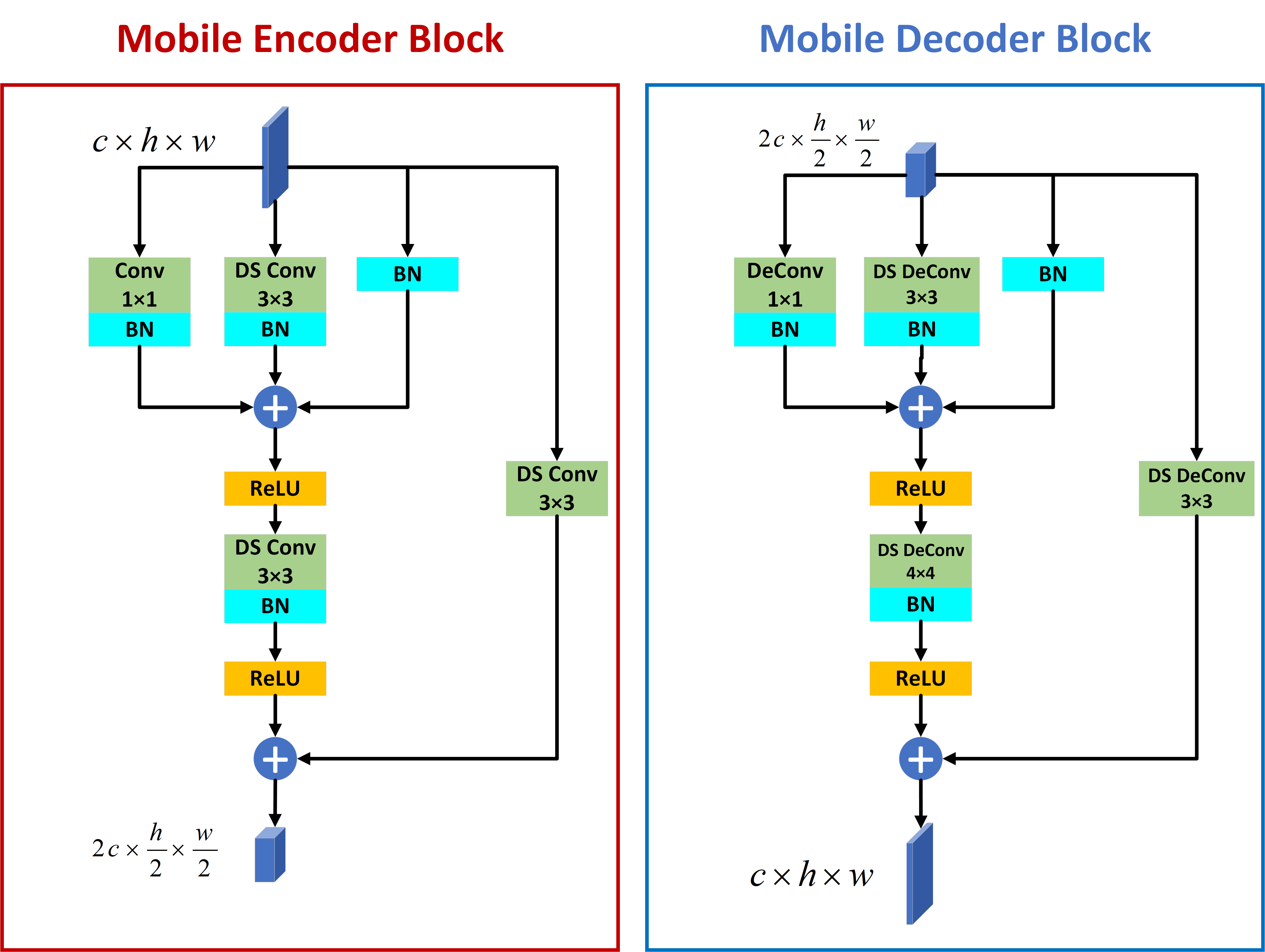}
\caption{The structure of mobile encoder block and mobile decoder block.}
\label{encoder_decoder}
\end{figure}

Convolutional Autoencoder (CAE) \cite{de2020deep} and other convolution-based denoising autoencoder \cite{rock2019cnns}\cite{rock2019complex} all adopt the combination of normal convolution (de-convolution) and max-pooling. From our perspective, this combination has two drawbacks. Firstly, the normal convolution (de-convolution) has many redundant parameters, which are likely to represent the noise signals. Secondly, max-pooling leads to information loss seriously. 

Based on the above, we design a novel and effective combination of encoder and decoder blocks in Radar-STDA, which are mobile encoder (decoder) block and efficient channel attention (ECA) block.

Inspired by MobileOne \cite{vasu2022improved}, we design a low-cost and high-performance encoder block called mobile encoder block. In the mobile encoder block, the feature map $x_i \in R^{c \times h \times w}$ will be firstly processed by three different branches. The first branch $\hat{x_i}^1$ is the convolution with $1 \times 1$ kernel size ($Conv_{1\times1}$) and a batch normalization operation ($BN$). The second branch $\hat{x_i}^2$ is the convolution with $3 \times 3$ kernel size ($Conv_{3\times3}$) and a batch normalization operation. The third branch $\hat{x_i}^3$ is a batch normalization operation, which could be seen as a short residual side. After that, the outputs of these three branches will be element-wisely added. After that, the feature map $x_{i+1}$ will be activated by $ReLU$ \cite{agarap2018deep}. The whole process is shown in Eq. \ref{mobile_encoder_stage1}.

\begin{align}
&    \hat{x_i}^1 = BN(Conv_{1\times1}(x_i)), \hat{x_i}^1 \in R^{c \times h \times w}, \nonumber \\
&    \hat{x_i}^2 = BN(Conv_{3\times3}(x_i)), \hat{x_i}^2 \in R^{c \times h \times w}, \nonumber \\
&    \hat{x_i}^3 = BN(x_i), \hat{x_i}^3 \in R^{c \times h \times w}, \nonumber \\
&    x_{i+1} = \hat{x_i}^1 + \hat{x_i}^2 + \hat{x_i}^3, x_{i+1} \in R^{c \times h \times w}, \nonumber \\
&    x_{i+1} = ReLU(x_{i+1}), x_{i+1} \in R^{c \times h \times w}.
\label{mobile_encoder_stage1}
\end{align}

After that, the feature map $x_{i+1}$ will go through a convolution operation with $3 \times 3$ kernel size and 
stride step of 2 ($Conv_{3\times3-2}$), expanding feature dimensions and enlarging receptive fields. Then, a batch normalization operation ($BN$) and a $ReLU$ \cite{agarap2018deep} follow. Finally, the feature map will be added with a convolution operation with $3 \times 3$ kernel size and 
stride step of 2 ($Conv_{3\times3-2}$) in a long residual side from the original feature map $x_i$. The long residual could effectively alleviate the gradient explosion and vanishing. The whole process is shown in Eq. \ref{mobile_encoder_stage2}.

\begin{align}
&    \hat{x_{i+1}} = BN(Conv_{3\times3-2}(x_{i+1})), \hat{x_{i+1}} \in R^{2c \times \frac{h}{2} \times \frac{w}{2}}, \nonumber \\
&    x_{i+2} = ReLU(\hat{x_{i+1}}) + Conv_{3\times3-2}(x_i), x_{i+2} \in R^{2c \times \frac{h}{2} \times \frac{w}{2}},
\label{mobile_encoder_stage2}
\end{align}
where $x_{i+2} \in R^{2c \times \frac{h}{2} \times \frac{w}{2}}$ is the final output of a mobile encoder block.

The mobile decoder block has the same architecture as the mobile encoder block. The only difference is that all the convolution operations are replaced with de-convolution operations ($DeConv$). Assuming the input feature map of the mobile decoder block is $x_j \in R^{2c \times \frac{h}{2} \times \frac{w}{2}}$. The mobile decoder block could also be divided into 2 stages, which are shown in Eq. \ref{mobile_decoder_stage1} and Eq. \ref{mobile_decoder_stage2}.

\begin{align}
&    \hat{x_j}^1 = BN(DeConv_{1\times1}(x_j)), \hat{x_j}^1 \in R^{2c \times \frac{h}{2} \times \frac{w}{2}}, \nonumber \\
&    \hat{x_j}^2 = BN(DeConv_{3\times3}(x_j)), \hat{x_j}^2 \in R^{2c \times \frac{h}{2} \times \frac{w}{2}}, \nonumber \\
&    \hat{x_j}^3 = BN(x_j), \hat{x_j}^3 \in R^{2c \times \frac{h}{2} \times \frac{w}{2}}, \nonumber \\
&    x_{j+1} = \hat{x_j}^1 + \hat{x_j}^2 + \hat{x_j}^3, x_{j+1} \in R^{2c \times \frac{h}{2} \times \frac{w}{2}}, \nonumber \\
&    x_{j+1} = ReLU(x_{j+1}), x_{j+1} \in R^{2c \times \frac{h}{2} \times \frac{w}{2}}.
\label{mobile_decoder_stage1}
\end{align}

\begin{align}
&    \hat{x_{j+1}} = BN(DeConv_{4\times4-2}(x_{j+1})), \hat{x_{j+1}} \in R^{c \times h \times w}, \nonumber \\
&    x_{j+2} = ReLU(\hat{x_{j+1}}) + Conv_{3\times3-2}(x_j), x_{j+2} \in R^{c \times h \times w}.
\label{mobile_decoder_stage2}
\end{align}

\begin{figure}
\centering
\includegraphics[width=0.47\textwidth]{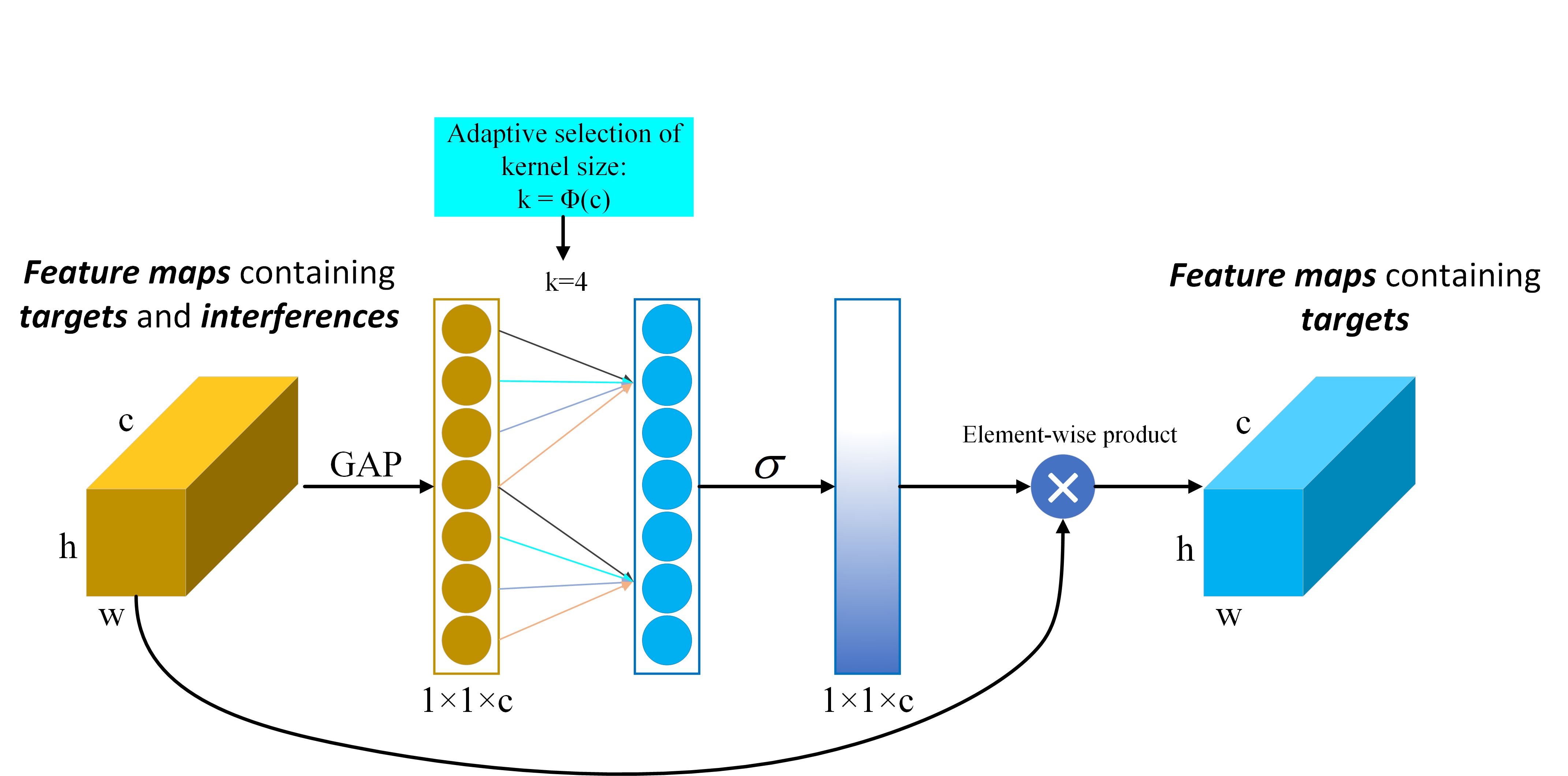}
\caption{The efficient channel attention (ECA) block to remove interferences on range-Doppler maps adaptively.}
\label{eca_block}
\end{figure}

However, due to the weight sharing of convolution kernels, both targets and interferences share common convolution kernels, which is not reasonable. For a denoising autoencoder, only the mobile encoder (decoder) block is not enough, because there are many feature maps containing noises in the forward propagation. Therefore, the autoencoder must selectively attach importance to the feature maps with the information of the target and ignore the feature maps with interferences as much as possible. Based on these, we tactfully adopt efficient channel attention (ECA) to help to encode and decode the feature maps adaptively. As Fig. \ref{eca_block} presents, feature map $m \in R^{c \times h \times w}$ containing both targets and interferences will be firstly processed by a global average pooling operation (GAP), which gets $\tilde{m} \in R^{1 \times 1 \times C}$ as a result. Then $\tilde{m} \in R^{1 \times 1 \times C}$ would be processed by a 1-D convolution with a sigmoid function, which calculates the weight of each channel of $m \in R^{c \times h \times w}$ and we call the weighted channel matrix as $W_m \in R^{1 \times 1 \times C}$. Finally, feature map $m \in R^{c \times h \times w}$ and weighted channel matrix $W_m$ are multiplied in the format of an element-wise product. The result is the weighted-channel feature map $\hat{m} \in R^{c \times h \times w}$. The whole process is shown in Eq. \ref{eca_process_equation}.

\begin{align}
&    \tilde{m} = GAP(m), \tilde{m} \in R^{1 \times 1 \times C}, \nonumber \\
&    W_m = \sigma(Conv_{1D}(\tilde{m})), W_m \in R^{1 \times 1 \times C},\nonumber \\
&    \hat{m} = W_m m, \hat{m} \in R^{c \times h \times w},
\label{eca_process_equation}
\end{align}
where $\sigma$ represents the sigmoid function.

\subsection{Skip Connection}

Between the modules of the encoder and decoder, there are two skip connections, where the feature map in the decoder module will be added by the feature map of the same size in the encoder module. The skip connection firstly could dramatically alleviate the gradient vanishing as the model training goes on. Secondly, skip connections could combine multi-scale features in the model. It could make the decoder find the correct optimizing direction in the decoding process with the assistance of the encoder.

\section{Experimental setup}

As a result of the difficulties in acquiring both interfered radar echoes and the related references in practice, especially for dynamic scenarios, there are no available real-world data sets for interference suppression tasks. However, the model trained with simulated data sets may not perform well in a practical application for the lack of clutters and background noise in realistic environments. In this paper, we decide to synthesize real-world data with simulated interference signals for the proposed neural network training and then employ both synthetic and measured data for test.

\subsection {Synthetic data set}

\begin{figure}[ht!]
\centering
\includegraphics[width=3.4in]{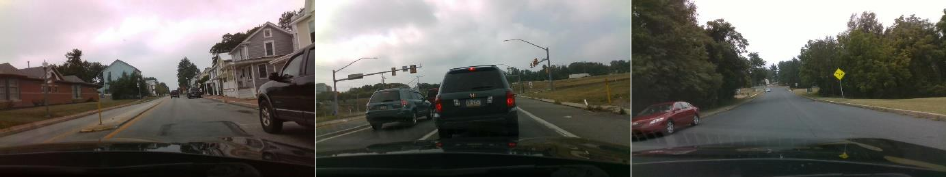}
\caption{Three example images corresponding to radar signals from the data set of RaDICaL.}
\label{Example images}
\end{figure}

For data generation, we consider a subset of RaDICaL\cite{lim2021radical} as victim signals, which gives a comprehensive list of radar configuration parameters described in Table \ref{tab:victim parameters}. The subset was collected from a mmWave radar inside a driving car, a model of AWR1843 BOOST from Texas Instrument. The roads driven are a myriad of neighborhood, suburban, highways and city roads. The views show oncoming traffic, incoming obstacles, street signs, guardrails when present, as well as the reflections from the car's engine and hood as shown in Fig. \ref{Example images}.

\begin{table}[]
    \centering
    \caption{Parameters of the victim radar that are collected from the open-source code of RaDICaL}
    \begin{tabular}{cccc}
    \hline
        Parameter & Value & Parameter & Value \\
        \hline
        Carrier frequency [GHz] & 77  & Sweep duration [$\mu s$] & 21.12 \\
        Sweep bandwidth [MHz] & 153.6  & Sampling frequency [MHz] & 12.5  \\
        Frame rate [frames/s] & 30 & ADC sampling window [$\mu s$] & 5.12  \\
        Max range [m] & 62.45 & Range resolution [m] & 0.97 \\
        Max velocity [m/s] & 23.02  & Velocity resolution [m/s] & 0.36 \\
        Samples per chirp & 64 & Chirps per frame & 128 \\
        Range FFT points & 64 & Doppler FFT points & 128 \\
        \hline
    \end{tabular}
    \label{tab:victim parameters}
\end{table}

\begin{table}[]
    \centering
    \caption{Parameters of the interfering radar used to generate interference signals}
    \begin{tabular}{cccc}
    \hline
        Parameter & Minimum & Maximum & Step \\
        \hline
        SINR [dB] & -5 & 25  & 5 \\
        Carrier frequency [GHz] & 76.8 & 77.2 & 0.1 \\
        Sweep bandwidth [MHz] & 120 & 400 & - \\
        Sweep duration [$\mu s$] & 4 & 30 & - \\
        Interferer distance [m] & 2 & 63 & - \\
        Interferer velocity [m/s] & -23.05 & 0 & -\\
        \hline
    \end{tabular}
    \label{tab:interferer parameters}
\end{table}

To emulate various scenarios with a single interference source, a few parameters, such as the sweep bandwidth and the interferer distance, are selected from intervals of uniform distribution by Monte-Carlo simulation. The detailed intervals for each parameter for interfering signals are listed in Table \ref{tab:interferer parameters}. Moreover, the interference amplitude is scaled by the Signal to Interference and Noise Ratio (SINR) from -5 dB to 25 dB with a fixed step of 5 dB. The simulated interference signal is then transmitted by an FMCW sensor via MATLAB R2021b Radar Toolbox and superimposed to the victim signal according to \ref{IF signal}. After synthesis of the time-domain data, the RD map of the interfered signal is computed by a 2D-FFT with a dimension of 64 $\times$ 128. Similarly, a 2D-FFT is implemented on the original signals, leading to the RD map of the associated reference. Since the strong clutter from the engine and hood of the car presents as ghosts in the RD maps around zero range and velocity axis, we filter them out by thresholding. 

Totally, there are six sequences in the subset, comprising 54,967 frames. Each frame is then combined with seven interference magnitudes on the basis of the various values of SINR mentioned above. The augmented data set contains 384,769 frames, which are standardized and normalized. To be trained by our Radar-STDA, three consecutive frames constitute a sample. Lastly, we have a data set of 128,247 samples and randomly split it into three partitions for training (60 $\%$), validation (20 $\%$) and test (20 $\%$). 

In contrast to the purely simulated data sets, the synthetic data set covers complex environment, such as the clutter from static buildings and the radar sensor itself. On the other hand, the measurement signals without interference can be regarded as references, which are barely accessible by labeling the interfered measurement.

\subsection {Performance metrics}

For performance evaluation, we examine the results with both quantitative and qualitative metrics. One of these quantitative measures is the SINR, which is defined by the ratio of the average power at object peaks to the noise floor \cite{toth2019performance}. In a 2-dimensional RD map, the SINR is computed as

\begin{equation}
    SINR=10\log({\frac{\frac{1}{N_O}\sum_{\{n,m\}\in O}{|\Tilde{S}_{RD}[n,m]|^2}}{\frac{1}{N_N}\sum_{\{n,m\} \in N}{|\Tilde{S}_{RD}[n,m]|^2}}}),
\end{equation}
where $n$ and $m$ are row and column indices of the RD matrix, $O$ is the set of target peaks and $N$ is the set of noise cells. The error vector magnitude (EVM) is another quantitative indicator, defined as the magnitude of the error vector between the clean RD map $S_{RD,clean} $and the noise-degraded RD map $\Tilde{S}_{RD}$ \cite{toth2019performance}:

\begin{equation}
    EVM=\frac{1}{N_O} \sum_{\{n,m\}\in O}{\frac{|S_{RD,clean}[n,m]-\Tilde{S}_{RD}[n,m]|}{|S_{RD,clean}[n,m]|}}.
\end{equation}

While the SINR gives information about the detection probability, the EVM measures the distortion in detected object properties, i.e. the object peak's magnitude and phase. Thus the goal of interference suppression is to maximize SINR and minimize EVM. 

Thirdly, as the Cell Averaging-Constant False Alarm Rate (CA-CFAR) \cite{1991Statistical} is applied as a peak detector on the denoised RD maps, we use Average Precision (AP) to measure the ratio of correctly detected objects to the total number of peaks. Mathematically, the AP can be defined as

\begin{equation}
    AP=\frac{N_{cd}}{N} \times 100\%,
\end{equation}
where $N_{cd}$ is the number of correctly detected peaks and $N$ is the number of total peaks in the reference.

Besides, the visual inspection of the RD map is considered a method of qualitative measurement, through observing object peak and noise floor intensity, as well as object resolution and peak distortion.

\subsection{Training settings of Radar-STDA}

\begin{table}[]
    \centering
    \caption{Training settings of Radar-STDA}
    \begin{tabular}{cc}
    \hline
        Hyperparameters & Values \\
        \hline
        batch size & 16 \\
        epoch & 100 \\
        initial learning rate & 0.001 \\
        optimizer & AdamW \cite{loshchilov2017decoupled} \\
        weight decay & 5e-4 \\
        scheduler & step annealing \\  
        \hline
    \end{tabular}
    \label{train_settings_radar_stda}
\end{table}

As TABLE \ref{train_settings_radar_stda} presents, we train our Radar-STDA on two RTX A4000 GPU with batch size 16. It is trained for 100 epochs with an initial learning rate of 0.01. We use AdamW \cite{loshchilov2017decoupled} as the optimizer with a weight decay of 5e-4 and step annealing as the scheduler.

\section{Results}

For a comparative analysis of the proposed autoencoder-based architecture, two classical methods are selected: Zeroing and IMAT. Additionally, we regard multi-layer perception (MLP) and convolutional autoencoder (CAE) as DL-based comparison methods.

\subsection{Quantitative evaluations}

Table \ref{tab:metrics} indicates the value of each metric for different methods, where Labels are the raw signals without interference. It can be seen that our Radar-STDA shows a better performance compared with other methods. Regarding the traditional methods, the SINR of our network is 7.34 $\%$ and 5.01 $\%$ higher than that of Zeroing and IMAT respectively. Besides, our Radar-STDA outperforms the two DL-based methods by 4.14 $\%$ and 1.36 $\%$ in terms of SINR. 

Since EVM measures the similarity between two objects in ground truth and inference, a lower value implies a more minor difference. The EVM values in Table \ref{tab:metrics} indicate that our method achieves the best results with 0.1792, followed by the two DL-based methods, due to the excellent effect of attention mechanism and temporal information on the task. For the AP calculation, CA-CFAR is used to detect the peaks in the RD maps and then these peaks are clustered by DBSCAN \cite{ester1996density}. Furthermore, the targets in RD maps are labeled by the annotation generator in \cite{ouaknine2021carrada}. The average precision is 87.15 $\%$ on Radar-STDA which reaches the best results of all the methods. In conclusion, Radar-STDA surpasses the most advanced algorithms in respect of interference mitigation.

\begin{table}[]
    \centering
    \caption{Performance comparison with the state-of-the-art studies on interference mitigation}
    \begin{tabular}{cccc}
    \hline
        Method & SINR [dB] & EVM & AP [\%] \\
        \hline
        Ground Truth & 18.28 & 0 & 89.17 \\
        Zeroing & 9.74 & 0.3994 & 58.29 \\
        IMAT & 12.07 & 0.3251 & 73.84 \\
        AECNet(MLP) & 12.94 & 0.3128 & 75.79\\
        CAE \cite{rock2019complex} & 15.72 & 0.2247 & 83.23\\
        Radar-STDA & 17.08 & 0.1792 & 87.15 \\
        \hline
    \end{tabular}
    \label{tab:metrics}
\end{table}


\subsection{Qualitive evaluations}

\begin{figure}
\centering
\includegraphics[width=0.47\textwidth]{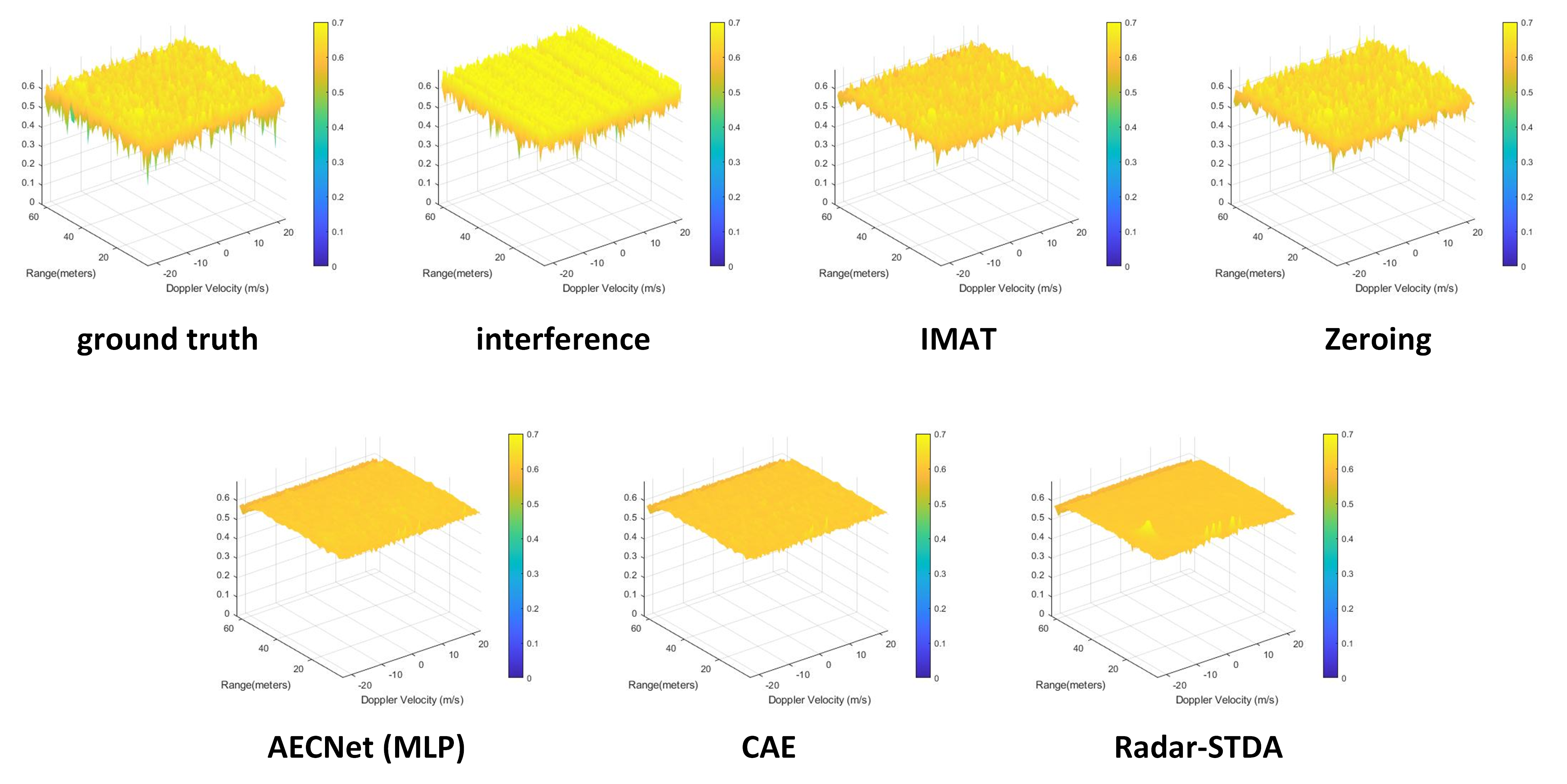}
\caption{Denoising results of five different denoising methods.}
\label{matlab_visual}
\end{figure}

Fig. \ref{matlab_visual} presents the range-Doppler maps of ground-truth, interference and five denoising methods. For signal-processing-based methods, IMAT outperforms Zeroing, which makes the target peak more apparent. For DL-based methods, we could see each model could remove the noise and make the plane of range-Doppler map flat and smooth. However, AECNet (MLP) and CAE \cite{rock2019complex} mistakenly remove the target peak along with the interference. In contrast, Radar-STDA maintains all target peaks successfully.

\begin{figure}
\centering
\includegraphics[width=0.47\textwidth]{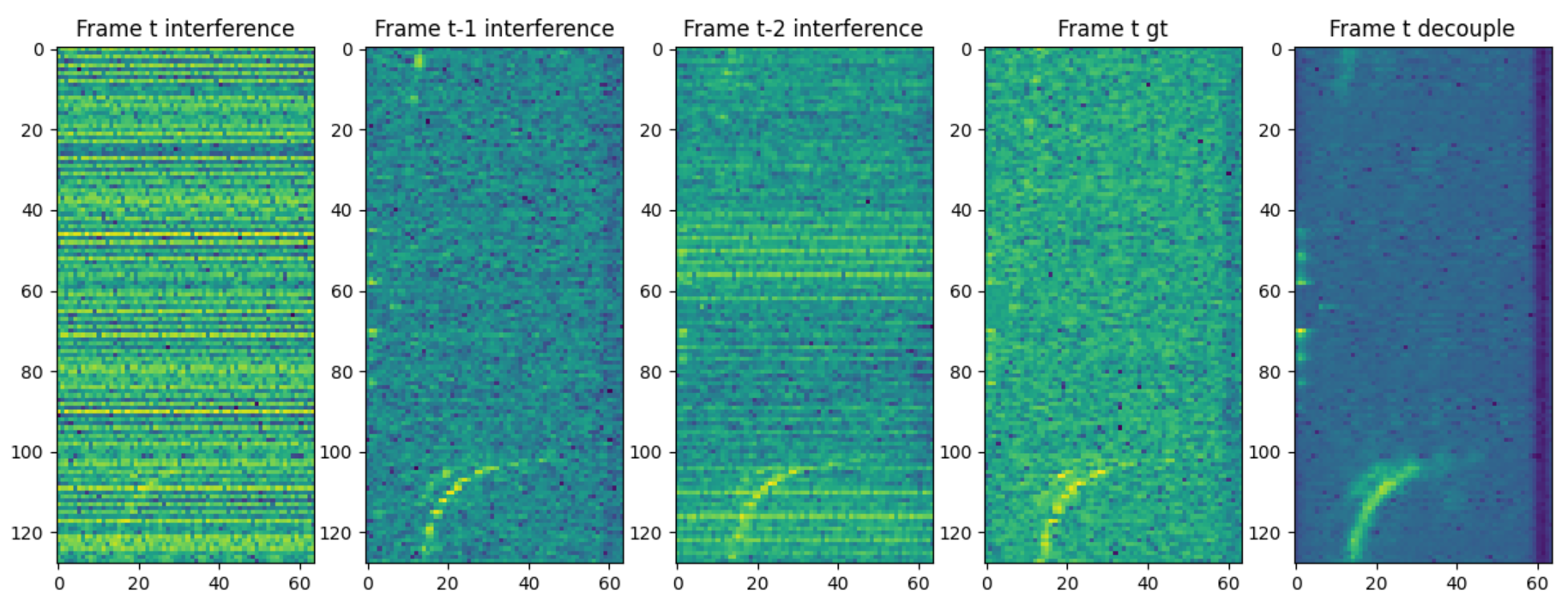}
\includegraphics[width=0.47\textwidth]{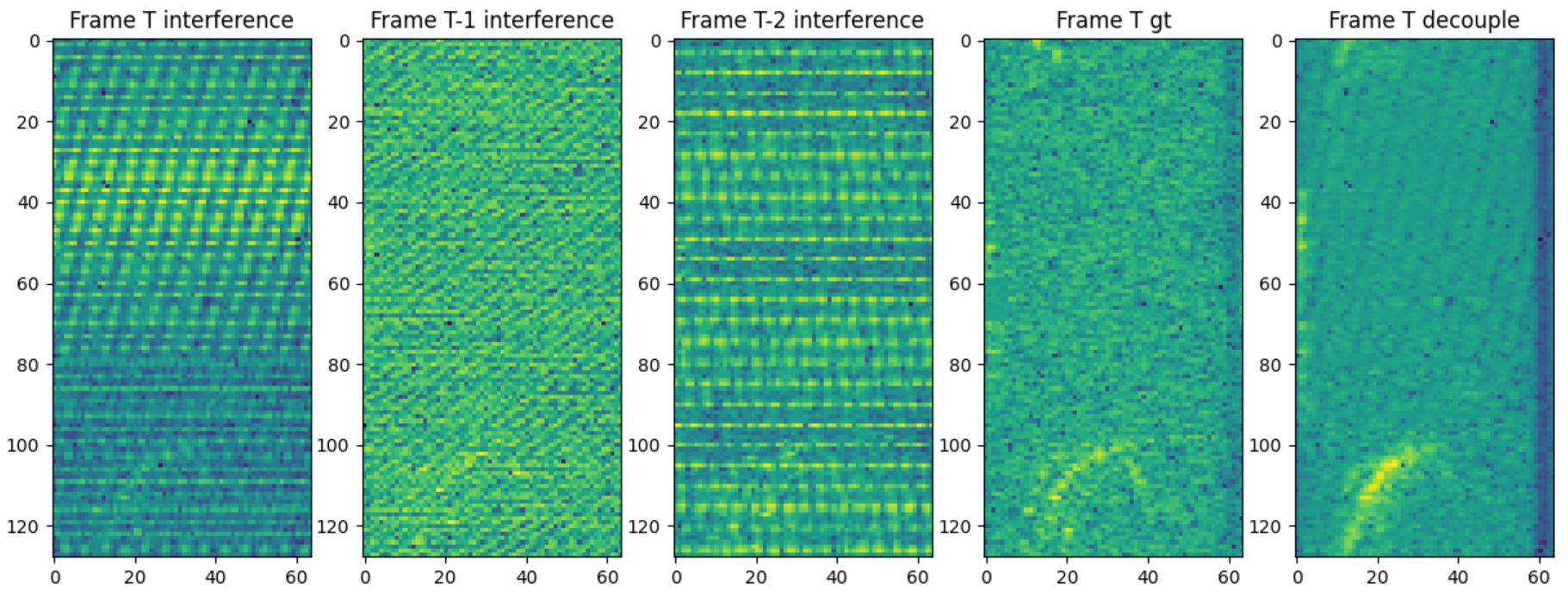}
\includegraphics[width=0.47\textwidth]{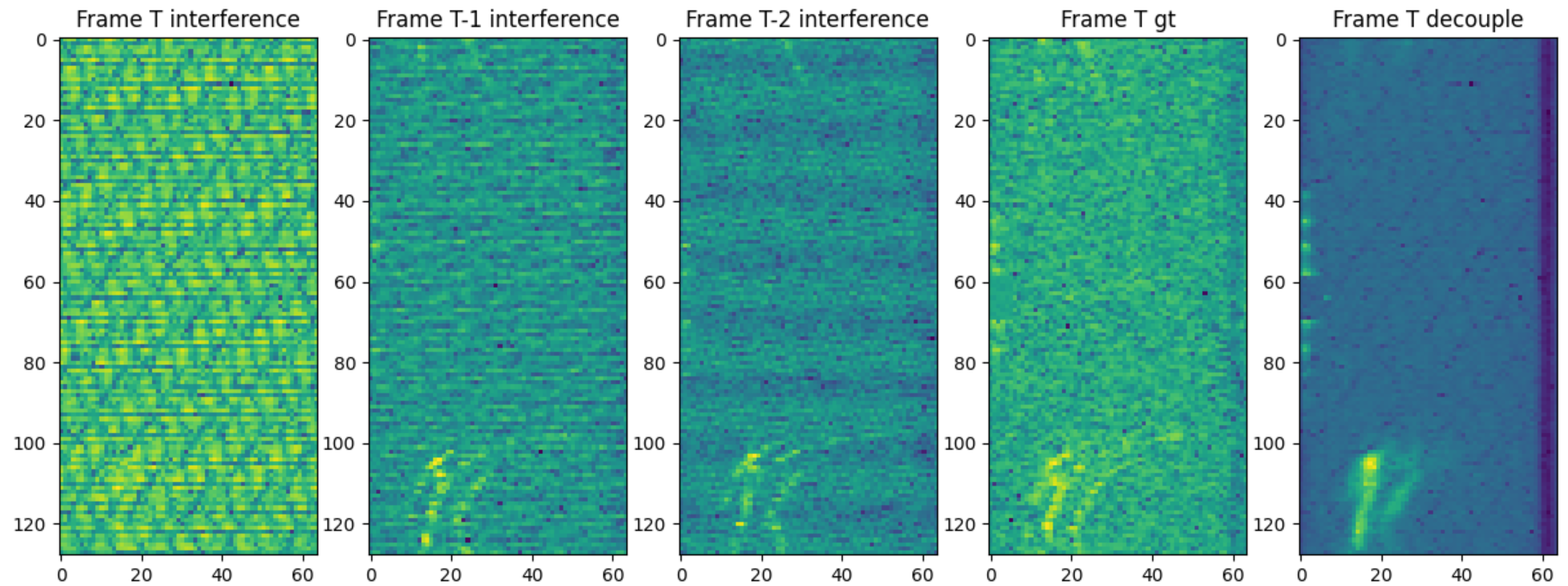}
\includegraphics[width=0.47\textwidth]{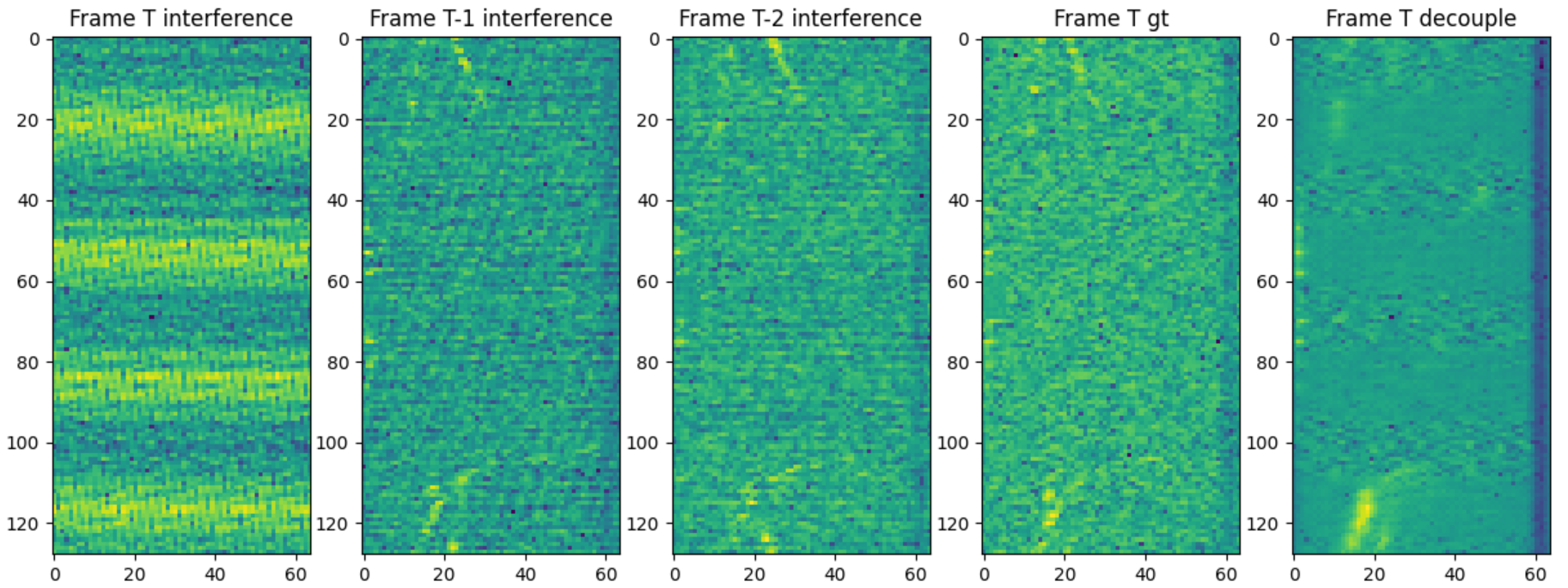}
\caption{Denoising results of Radar-STDA. The visualization is through the function $imshow()$ of OpenCV. The first, second and third columns list the range-Doppler maps at frame $T$, $T-1$ and $T-2$ with interference. The fourth column lists the ground-truth of range-Doppler map without interference. The last column lists the denoised range-Doppler maps by Radar-STDA.}
\label{python_visual}
\end{figure}

Fig. \ref{python_visual} shows the interference mitigation results of Radar-STDA. As we can see, the range-Doppler map at frame $T$ (the first column) has low SINR and the target peak could not be recognized. By fusing the range-Doppler maps at frames $T-1$ and $T-2$ (the second and third columns), the target peaks in denoised range-Doppler maps by Radar-STDA could be restored and easily recognized (the last column). Moreover, the fusion of features at different frames makes Radar-STDA need to address interference and noise of three frames in training, which are much more than that of frame $T$. It indirectly enhances the ability of Radar-STDA to mitigate interference and original noise.

\subsection{Computational efficiency}

\begin{table*}[]
    \centering
    \caption{Performances of DL-based models on different devices}
    \begin{tabular}{cccccc}
    \hline
        Models & Parameters (M) & MACs (M) & FPS  & FPS  & FPS  \\
               &                &          &(RTX A4000) & (i7-12700) &  (Nvidia Jetson AGX Xavier) \\
        \hline
        Radar-STDA & 0.14 & 101.37 & 207.57 & 31.67 & 56.82 \\
        CAE \cite{rock2019complex} & 5.85 &  $1.82\times10^4$ & 238.42 & 24.19 & 48.63 \\
        AECNet (MLP) & 1.35 & 171.97 & 1715.11 & 582.70 & 69.73 \\
        \hline
    \end{tabular}
    \label{performance_radar_stda}
\end{table*}

As TABLE \ref{performance_radar_stda} presents, compared with CAE \cite{rock2019complex} and AECNet(MLP), our Radar-STDA has only 0.14 million parameters, which takes up little memory during interference. Radar-STDA reaches 207.57 FPS to denoise one range-Doppler map on one RTX A4000 GPU, which is slower than CAE (238.42 FPS) and AECNet (MLP) (1715.11 FPS), because Radar-STDA takes three range-Doppler maps as the parallel input. In contrast, CAE \cite{rock2019complex} and AECNet (MLP) adopt only one range-Doppler map as the input, which would be faster on the high-performance GPU device. Due to full depthwise separable convolution operation that Radar-STDA adopts, it could run faster than models with normal convolution where the hashrate of hardware devices are limited. Therefore, it could be observed that Radar-STDA run faster than CAE \cite{rock2019complex} on one i7-12700 CPU and NVIDIA Jetson AGX Xavier (edge device). Radar-STDA gets 31.67 and 56.82 FPS on one i7-12700 CPU and NVIDIA Jetson AGX Xavier respectively. By contrast, CAE \cite{rock2019complex} achieves 24.19 and 48.63 FPS on two devices. Besides, AECNet, as a full MLP architecture model, has the fastest inference speed on three different devices. However, the performance of AECNet (MLP) on interference mitigation is worse than CAE \cite{rock2019complex} and Radar-STDA (TABLE \ref{tab:metrics}). All in all, Radar-STDA has a nano-level parameter size and could infer in real-time on different devices.

\section{Conclusion}

In this paper, we investigate the impact of interference between FMCW radar sensors that 'ridge', 'ghost target' and 'increased noise floor' may be due to different chirp configurations. For the task research, the synthetic data set called Ra-inf is released in order to obtain an environment closer to reality. Moreover, we propose a DL-based denoisng autoencoder called Radar-STDA. It utilizes the attention mechanism and the fusion of spatial-temporal information that outperforms the state-of-the-art methods, such as Zeroing, IMAT and CAE for interference mitigation and denoising. Our network reaches 17.08 dB in SINR which means that it can be restored remarkably close to the ground truth. Meanwhile, our model obtains the minimum distortion of all methods with an EVM of 0.1792. The accuracy of the target detection after interference mitigation by the proposed method is 87.15 $\%$ and is also the best of all methods. More importantly, Radar-STDA has a nano-level size and allows for real-time denoising of range-Doppler maps on both host and edge devices. In summary, the proposed approach achieves state-of-the-art performance in various metrics compared with the well-known traditional and DL-based techniques. In future work, we aim to design an integrated neural network for interference suppression while detecting, which could considerably reduce the computational cost.

my reference \cite{zhao2021point}, \cite{ma2022rethinking}, \cite{qian2022pointnext}, \cite{buhmann2000radial}, \cite{he2016deep}, \cite{schumann2021radarscenes}

%


\section*{Declaration}
Lulu Liu and Runwei Guan contribute equally to this paper.

\section*{Acknowledgment}
This work received financial support from Jiangsu Industrial Technology Research Institute(JITRI) and Wuxi National Hi-Tech District(WND).

\ifCLASSOPTIONcaptionsoff
  \newpage
\fi



%
\bibliographystyle{IEEEtran}
\bibliography{IEEEabrv,Bibliography}

%

\begin{IEEEbiography}[{\includegraphics[width=1in,height=1.25in,clip,keepaspectratio]{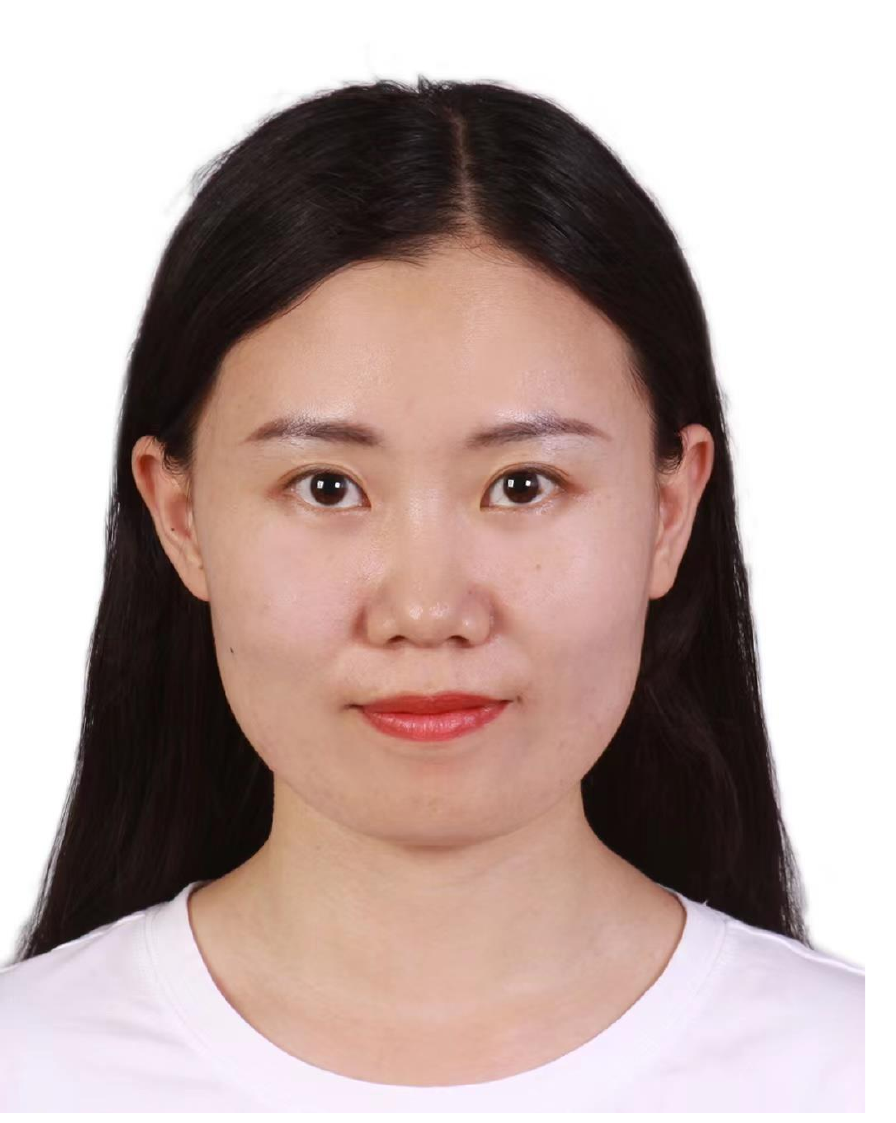}}]{Lulu Liu}
received the M.Sc. degree in information and communication engineering from University of Science and Technology of China, Hefei, China, in 2017. She is currently working toward the Ph.D. degree with Xi'an Jiaotong-Liverpool University in cooperation with the Institute of Deep Perception Technology, Jiangsu Industrial Technology Research Institute. Her research interests include signal processing and deep learning with a focus on automotive radar, interference mitigation and target detection.
\end{IEEEbiography}

\begin{IEEEbiography}[{\includegraphics[width=1in,height=1.25in,clip,keepaspectratio]{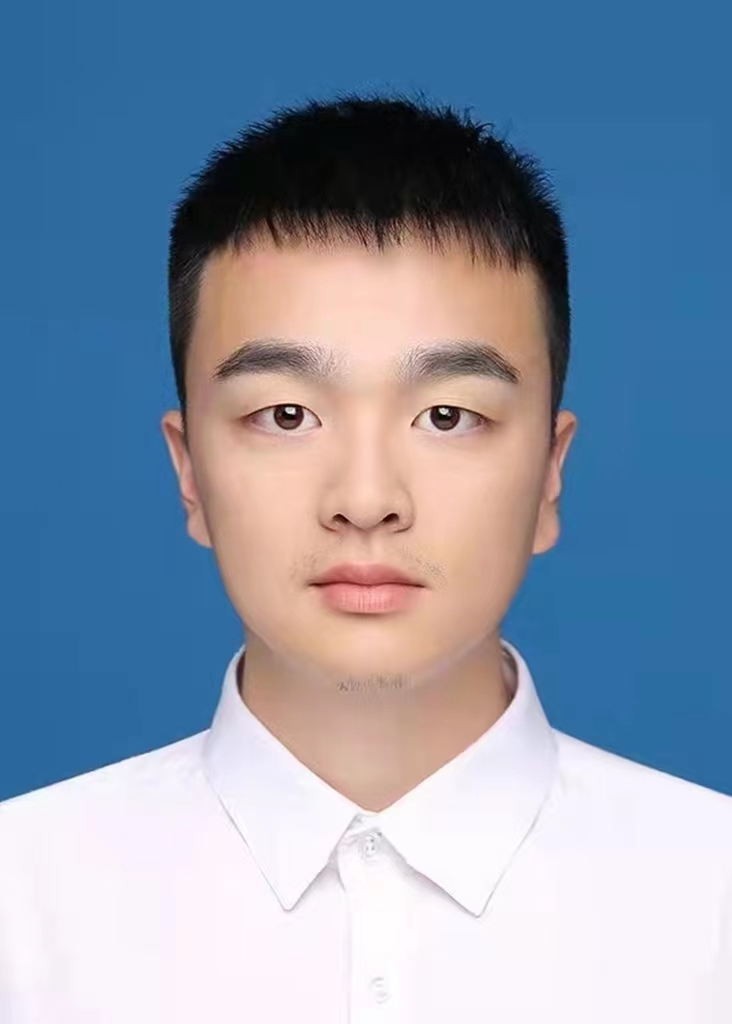}}]{Runwei Guan}
(Student Member, IEEE), received his M.S. degree in Data Science from University of Southampton, Southampton, United Kingdom, in 2021. He is currently a joint Ph.D. student of University of Liverpool, Xi'an Jiaotong-Liverpool University and Institute of Deep Perception Technology, Jiangsu Industrial Technology Research Institute. His research interests include multi-modal learning based on computer vision and natural language processing, object detection based on the fusion of millimeter-wave radar and camera, lightweight neural network, vision transformer and statistical machine learning.
\end{IEEEbiography}


\begin{IEEEbiography}[{\includegraphics[width=1in,height=1.25in,clip,keepaspectratio]{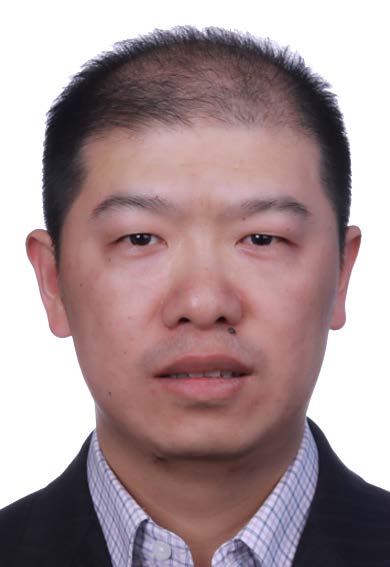}}]{Fei Ma}
holds a PhD in Applied Mathematics from the Flinders University of Australia. He received his B.Sc and M.Sc. degrees of Computational Mathematics from Xiamen University, China. Upon joining Xi'an Jiaotong-Liverpool University, he has been an Analyst of Symbion, Australia, Research associate and lecturer of Flinders University, and Software Engineer of Kingdee Co. Ltd in Shenzhen. He is currently a professor at the Applied Mathematics Department of Xi'an Jiaotong-Liverpool University, Suzhou, China.Dr. Ma’s research interests include: medical and biomedical image analysis; big data analytics; inventory forecasting; graph matching; and numerical algebra.

\end{IEEEbiography}

\begin{IEEEbiography}[{\includegraphics[width=1in,height=1.25in,clip,keepaspectratio]{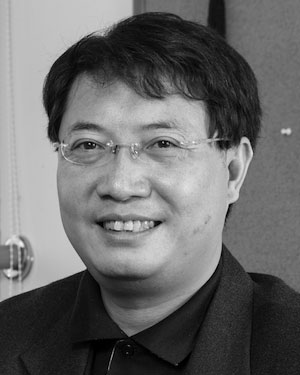}}]{Ka Lok Man}
received the Dr.Eng. degree in electronic engineering from the Politecnico di Torino, Turin, Italy, in 1998, and the Ph.D. degree in computer science from Technische Universiteit Eindhoven, Eindhoven, The Netherland, in 2006. He is currently a Professor in Computer Science and Software Engineering with Xi’an Jiaotong-Liverpool University, Suzhou, China. His research interests include formal methods and process algebras, embedded system design and testing, and photovoltaics.
\end{IEEEbiography}


\begin{IEEEbiography}[{\includegraphics[width=1in,height=1.25in,clip,keepaspectratio]{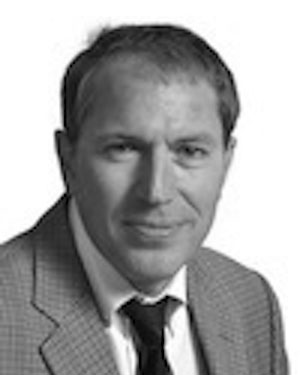}}]{Jeremy Smith}
(Member, IEEE) received the B.Eng. (Hons.) degree in engineering science and the Ph.D. degree in electrical engineering from the University of Liverpool, Liverpool, U.K., in 1984 and 1990, respectively. Between 1984 and 1988, he was conducting research on image processing and robotic systems in the Department of Electrical Engineering and Electronics, University of Liverpool, Liverpool, U.K., where he was a Lecturer, Senior Lecturer, and Reader in the same department since 1988. Since 2006, he has been a Professor in electrical engineering with the University of Liverpool. His research interests include automated welding, robotics, vision systems, adaptive control, and embedded computer systems.
\end{IEEEbiography}

\begin{IEEEbiography}[{\includegraphics[width=1in,height=1.25in,clip,keepaspectratio]{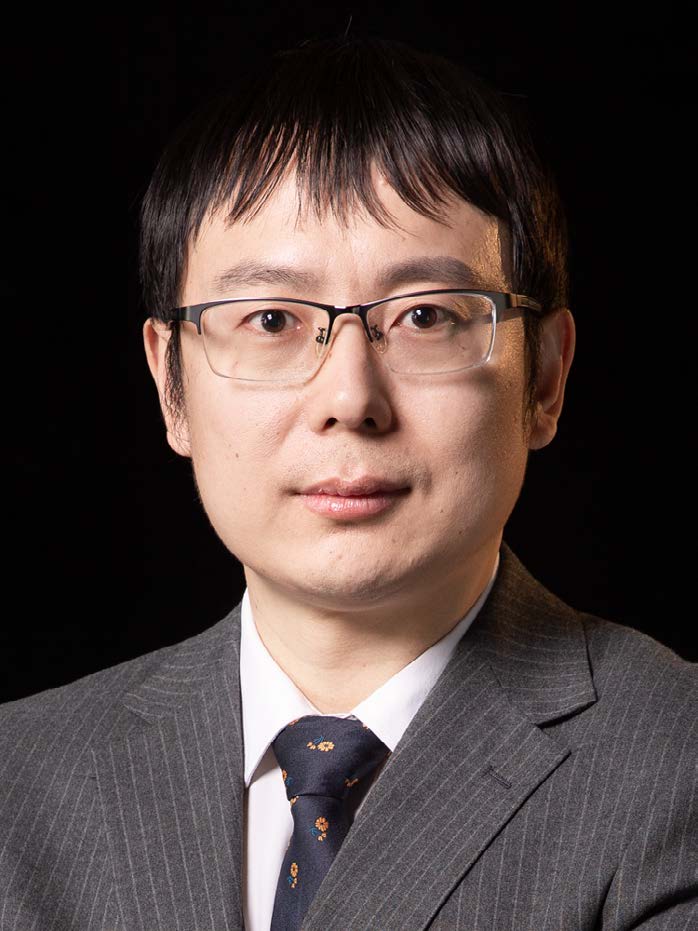}}]{Yutao Yue}
was born in Qingzhou, Shandong, China, in 1982. He received the B.S. degree in applied physics from the University of Science and Technology of China, in 2004, and the M.S. and Ph.D. degrees in computational physics from Purdue University, USA, in 2006 and 2010, respectively. From 2011 to 2017, he worked as a Senior Scientist with the Shenzhen Kuang-Chi Institute of Advanced Technology and a Team Leader of the Guangdong ``Zhujiang Plan'' 3rd Introduced Innovation Scientific Research Team. From 2017 to 2018, he was a Research Associate Professor with the Southern University of Science and Technology, China. Since 2018, he has been the Founder and the Director of the Institute of Deep Perception Technology, JITRI, Jiangsu, China. Since 2020, he has been working as an Honorary Recognized Ph.D. Advisor of the University of Liverpool, U.K., and Xi'an Jiaotong-Liverpool University, China. He is the co-inventor of over 300 granted patents of China, USA, and Europe. He is also the author of over 20 journals and conference papers. His research interests include computational modeling, radar vision fusion, perception and cognition cooperation, artificial intelligence theory, and electromagnetic field modulation. Dr. Yue was a recipient of the WuWen Jun Artificial Intelligence Science and Technology Award in 2020.
\end{IEEEbiography}




\end{document}